\documentclass[aps,pra,superscriptaddress,amsmath,amssymb,preprintnumbers,floatfix,showpacs,12pt]{revtex4}
\usepackage{amssymb}
\usepackage{epsfig}
\usepackage{subfigure}
\usepackage{graphicx}
\begin{document}
\title{ Analytical approach for treating unitary quantum systems with  initial mixed states }

\author{Faisal A. A. El-Orany }
\email{el_orany@hotmail.com, Tel:006-166206010, Fax:
0060386579404}\affiliation{ Department of Mathematics and Computer
Science, Faculty of Science, Suez Canal University, Ismailia,
Egypt; } \affiliation{ Cyberspace Security Laboratory, MIMOS
Berhad, Technology Park Malaysia, 57000 Kuala Lumpur, Malaysia}

\begin{abstract}
The mixed states  are important in quantum optics since they
frequently appear in the decoherence problems. When one of the
components of the system is prepared in the mixed state and the
evolution operator of this system is not available, one cannot
deduce the density matrix.
 We present analytical approach  to accurately solve this problem.
 The approach  can be applied on the condition that
 the Schr\"{o}dinger's equation of the system   is  solvable with
 any arbitrary initial state.
In deriving the solution we exploit the fact that any mixed state
can be expressed in terms of a phase state.
 The approach is illustrated by
 deriving the density matrix of  a single-mode heat
environment interacting asymmetrically with two qubits. Our
results  are in good agreement with the available  results in the
literature.  This approach opens new perspectives for treating
complicated systems and may impact other applications in the
quantum theory.

\end{abstract}
 \maketitle


\section{Introduction}

 In  quantum theory, the state of the system has two
categories; namely pure state and mixed state.  A pure state
implies perfect knowledge  of the system. For the mixed case, we
do not have enough information to specify the state of the system
completely and hence cannot form its wavefunction.  In this
situation, we can only describe the system via the density matrix
$\hat{\rho}_f$. Mixed state frequently appears in the decoherence
problems, e.g., \cite{bernet}. We can obtain the mixed state,
which is associated with any state, by considering its phase to be
totally randomized.  The best example of the mixed state is the
thermal field, which represents an electromagnetic radiation
emitted by a source at temperature $T$. The examination of a
quantum beam with a thermal noise is an important topic from both
theoretical and practical points of view, e.g. \cite{thermal}.

  In the  quantum information theory, a considerable
attention has been paid to the entanglement of the bipartite and
the multipartite systems in which one of the subsystems exists
initially in the thermal equilibrium \cite{amesen,lee1,lee2}. The
conclusion of all these studies is that it is possible for the
thermal field, which is a highly chaotic field, to induce
entanglement between qubits. The previous studies have been
limited to the systems whose exact form of the evolution operators
is obtainable \cite{amesen,lee2}. For the other systems the
solution  is difficult or even impossible. For some few systems of
the latter, one has to use the numerical methods to solve  the
master equation of the system \cite{lee1}, but the results cannot
be totally  trusted. In this paper we develop, for the first time,
a simple analytical approach  solving  this problem exactly. This
approach can provide
 the dynamical density matrix of the unitary quantum system
whose one of its components is initially in the mixed state such
as the thermal field. It works only when  the Schr\"{o}dinger's
equation of the system for any arbitrary initial state is
solvable. Furthermore, the approach can be applied as an
alternative technique for finding the solution of the systems
whose evolution operators exist but they are very complicated. For
instance, the evolution of the mixed field   with the  multi-level
atoms, e.g. three-level atom and four-level atom, etc.

In  section II we describe the approach  in details. In section
III we  give an  application for the approach by solving specific
problem. The example, which we have considered, is that the
problem of the two atoms in the cavity interact asymmetrically
with the thermal field. The reason behind this choice  is that
this system is  a subject of the current research in relation to
entanglement \cite{amesen,lee1,lee2}. Thus, we can easily validate
the approach by comparing our results with the available results
in the literature.

%%%%%%%%%%%%%%%%%%%%%%%%%%%%%%%%%%%%%%%%%%%%%%%%%%%%%%%%%%%%%%%%%
\section{Description of the approach}
%%%%%%%%%%%%%%%%%%%%%%%%%%%%%%%%%%%%%%%%%%%%%%%%%%%%%%%%%%%%%%%%%
In this section,  we describe how one  can deduce
 the  density matrix of the unitary quantum system
whose one of its components is initially prepared in the thermal
state. Before going into  details, let us briefly state some
properties of the thermal field. The thermal field has a diagonal
expansion
in terms of Fock states as \cite{bernet}: %%%%%%%%%%%%%%%%%%%%%%%%%%%%%%%%%%%%%%
\begin{equation}
\hat{\rho}_f=\sum\limits_{n=0}^{\infty}p(n)|n\rangle\langle
n|=\frac{1}{\pi}\int\limits_{0}^{\pi}|z(\phi)\rangle\langle
z(\phi)| d\phi, \label{1.2}
\end{equation}%
where $p(n)$ is the photon-number distribution of the thermal
field and $|z(\phi)\rangle$  is the associated  phase state having
the form:
\begin{equation}
|z(\phi)\rangle=\sum\limits_{n=0}^{\infty}\sqrt{p(n)}\exp(in\phi)|n\rangle
=\sum\limits_{n=0}^{\infty}C_n(\phi)|n\rangle,
 \label{1.3}
\end{equation}
where
\begin{equation}
C_n(\phi)=\sqrt{p(n)}\exp(in\phi),\quad
p(n)=\frac{\bar{n}^n}{(1+\bar{n})^{n+1}}
 \label{1.4}
\end{equation}
and $\bar{n}$ is the mean-photon number of the thermal field
having the form $\bar{n}=(e^{\hbar w/k_b T}-1)^{-1}$ and $k_b$ is
the Boltzmann's constant. It is obvious that  $\bar{n}$ increases
by increasing the temperature $T$.

Now we are in a position to explain the approach.
 Assume that we have a bipartite system whose Hamiltonian is
 $\hat{H}_{db}$, where $d$ and $b$ stand for the field and the other party,
 respectively. These two components are initially prepared in the
 states
 $|\psi_d(0)\rangle$ and  $|\psi_b(0)\rangle$.
The Schr\"{o}dinger's equation of this system is solvable and
hence we can obtain the wavefunction as $|\psi(t)\rangle$. From
the basic concepts   of the quantum mechanics we have:
\begin{equation}\label{ficn1}
   |\psi(t,0)\rangle=\hat{U}(t,0)|\psi_d(0)\rangle\bigotimes|\psi_b(0)\rangle,
\end{equation}
where $\hat{U}(t,0)$ is the evolution operator of the system
regardless if we can deduce its explicit form or not. Under these
assumptions we can solve this system accurately if the party $d$
is initially prepared in  the thermal field $\hat{\rho}_f$ instead
of  $|\psi_d(0)\rangle$. In this case the initial state of the
system reads:
\begin{equation}
\hat{\rho}(0)=\hat{\rho}_f\bigotimes |\psi_b(0)\rangle
\langle\psi_b(0)|=
\frac{1}{\pi}\int\limits_{0}^{\pi}|z(\phi)\rangle\langle
z(\phi)|\bigotimes|\psi_b(0)\rangle \langle\psi_b(0)| d\phi.
\label{ficn2}
\end{equation}%
Under the evolution operator $\hat{U}(t,0)$, the system in the
initial state (\ref{ficn2})  evolves as:
\begin{equation}
\hat{\rho}(t)=\hat{U}(t,0)\hat{\rho}(0)\hat{U}^{\dagger}(t,0)
=\frac{1}{\pi}\int\limits_{0}^{\pi}|\psi_z(t,\phi)\rangle\langle\psi_z
(t,\phi)| d\phi, \label{ficn3}
\end{equation}%
where
$|\psi_z(t,\phi)\rangle=\hat{U}(t,0)|z(\phi)\rangle\bigotimes|\psi_b(0)\rangle$.
Our target is to deduce the  wavefunction
$|\psi_z(t,\phi)\rangle$.  If the explicit form of $\hat{U}(t,0)$
is available, the solution is straightforward. If it is not, we
can solve the Schr\"{o}dinger's equation for the initial condition
$|z(\phi)\rangle$ and $|\psi_b(0)\rangle$.  We have to emphasize
if the system is solvable for the arbitrary state
$|\psi_d(0)\rangle$, it will be automatically solvable for the
state $|z(\phi)\rangle$. This is based on the fact that any state
of the field is just a linear combination of the Fock states
weighted by a specific  distribution. As soon as we derive
$|\psi_z(t,\phi)\rangle$ we substitute it into (\ref{ficn3}) and
carry out the integration over the phase $\phi$
 to obtain the requested density matrix.  From the above description, the theme of the approach
 we transform the problem of obtaining the dynamical density
matrix to that of finding the wavefunction of the system when the
 field  is initially prepared in the state $|z(\phi)\rangle$.

In spite of the simplicity of the approach, it is efficient and
able to provide  the exact solutions for some non-analytically
solvable problems so far. In the following section we show this
fact by  developing the analytical solution for one of those
problems, which has been  already numerically treated.
%%%%%%%%%%%%%%%%%%%%%%%%%%%%%%%%%%%%%%%%%%%%%%%%%
\section{Example:  two-qubit problem}
%%%%%%%%%%%%%%%%%%%%%%%%%%%%%%%%%%%%%%%%%%%%%%%%%
In this section,  we apply the approach described in the preceding
section to obtain the density matrix of the two-qubit problem.
Precisely, we deduce the dynamical density matrix of the system of
the single-mode thermal field interacting simultaneously with the
two two-level atoms in the cavity in the non-identical fashion.
The numerical technique has been exploited for solving this system
\cite{lee1}, however, here we present the analytical solution. It
is worthwhile mentioning that the two-identical-atom version of
this system has been already demonstrated, e.g.,
\cite{lee1,lee2,nf}. For the latter  the dynamical density matrix
of the system has been derived by the Kraus representation
\cite{kraus}. That cannot be applied to the non-identical-atom
system. Also, we verify the validity of  the approach  by
comparing some of our results to those  in the literature in
relation to entanglement.
  We proceed, under the rotating wave approximation the
Hamiltonian describing the system takes the form
\cite{lee1,lee2,nf}:
%%%%%%%%%%%%%%%%%%%%%%%%%%%%%%%%%%%%%%%%%%%%%%%%%%%%%%%%%%%%%%%%%%%%%%%%
\begin{eqnarray}
\begin{array}{lr}
\frac{\hat{H}}{\hbar}=\hat{H}_0+\hat{H}_I,\\
\\
\hat{H}_0= \omega\hat{a}^{\dagger}\hat{a}+
\frac{\omega_a}{2}(\hat{\sigma}_1^{z}+\hat{\sigma}_2^{z}),\quad
\hat{H}_I=\sum\limits_{j=1}^2 \lambda_j (\hat{a}\hat{\sigma}_j^{+}
+ \hat{a}^{\dagger }\hat{\sigma}_j^{-}),
 \label{6}
 \end{array}
\end{eqnarray}
%%%%%%%%%%%%%%%%%%%%%%%%%%%%%%%%%%%%%%%%%%%%%%%%%%%%%%%%%%%%%%%%%%%%%%%%%%%
where $\hat{H}_0$ and $\hat{H}_I$ are the free and interaction
parts of the Hamiltonian, $\hat{\sigma}_j^{\pm}$ and
$\hat{\sigma}_j^{z}$ are the Pauli spin operators of the $j$th
atom; $\hat{a}\quad (\hat{a}^{\dagger})$ is the annihilation
(creation) operator denoting  the cavity mode, $\omega$ and
$\omega_a$ are the frequencies of the cavity mode and the atomic
systems (we consider that the two atoms have the same frequency)
and $\lambda_j$ is the  coupling constant between the $j$th atom
and the field. Based on the relation between $\lambda_1$ and
$\lambda_2$ we have two cases; namely  asymmetric case
($\lambda_1\neq \lambda_2$) and symmetric case ($\lambda_1=
\lambda_2$). We should stress that  the asymmetric case is closer
to experiment than the symmetric one \cite{duan}. Finally,
throughout the investigation we assume that $\omega_a=\omega$.

Now,  our aim is to deduce the explicit form of the density matrix
$\hat{\rho}(t)$ of the Hamiltonian (\ref{6}) when the field is
initially prepared in the thermal field (\ref{1.2}) and the atoms
are initially
 in the following  mixed
state:
\begin{equation}
\hat{\rho}_a=\sin^2 \theta \cos^2 \vartheta|e_1,e_2\rangle\langle
e_2,e_1|+\sin^2 \theta \sin^2 \vartheta |g_1,g_2\rangle\langle
g_2,g_1| + \cos^2 \theta |e_1,g_2\rangle\langle g_2,e_1|,
\label{1.1}
\end{equation}%
where $|e_j\rangle$ and $|g_j\rangle $ stand for the excited and
the ground atomic states of the $j$th atom, respectively. The
variables $\theta$ and $\vartheta$ are phases, which can be
specified to give different types of the initial atomic states.
The subscript $a$ denotes the atomic system. We should stress that
the suggested approach is applicable for any initial atomic state
not only (\ref{1.1}).
  Noteworthily, for $\lambda_1\neq \lambda_2$ we cannot derive
the explicit form of the evolution operator of the system
$\hat{U}(t,0)$, however, we can solve the Schr\"{o}dinger's
equation  \cite{faisala}. The latter is the main demand  to apply
the approach.
 The whole initial state of the system can be easily written in the form (\ref{ficn2}).
Under the evolution operator $\hat{U}(t,0)$, this initial state
  evolves as:
\begin{eqnarray}
\begin{array}{lr}
\hat{\rho}(t)=\hat{U}(t,0)\hat{\rho}(0)\hat{U}^{\dagger}(t,0)\\
\\
=\frac{\sin^2 \theta \cos^2 \vartheta}{\pi}\int\limits_{0}^{\pi}
|\Psi_1(t,\phi)\rangle\langle \Psi_1(t,\phi)|d\phi +\frac{\sin^2
\theta \sin^2 \vartheta}{\pi}\int\limits_{0}^{\pi}
|\Psi_2(t,\phi)\rangle\langle \Psi_2(t,\phi)|d\phi\\
\\
+\frac{\cos^2\theta}{\pi}\int\limits_{0}^{\pi}
|\Psi_3(t,\phi)\rangle\langle \Psi_3(t,\phi)|d\phi,
 \label{1.7}
\end{array}
\end{eqnarray}
where
\begin{eqnarray}
\begin{array}{lr}
|\Psi_j(t,\phi)\rangle=\hat{U}(t,0)|\Psi_j(0,\phi)\rangle, \quad j=1,2,3,\\
\\
|\Psi_1(0,\phi)\rangle=|e_1,e_2,z(\phi)\rangle, \quad
|\Psi_2(0,\phi)\rangle=|g_1,g_2,z(\phi)\rangle, \quad
|\Psi_3(0,\phi)\rangle=|e_1,g_2,z(\phi)\rangle.
 \label{ficn5}
\end{array}
\end{eqnarray}
 Our main concern is to   derive
   the wave functions $|\Psi_j(t,\phi)\rangle$.  These can be easily obtained by  solving the Schr\"{o}dinger's equation
     three times for
the given initial  conditions  as \cite{faisala}: %%%%%%%%%%%%%%%%%%%%%%%%%%%%%%%%
\begin{widetext}
\begin{eqnarray}
\begin{array}{lr}
\mid \Psi_j (t,\phi)\rangle =\sum\limits_{n=0}^{\infty
}C_n(\phi)\left[ X_{1}^{(j)}(t,n)\mid e_{1},e_{2},n\rangle
+X_{2}^{(j)}(t,n)\mid
e_{1},g_{2},n+1\rangle \right. \\
\\
+\left. X_{3}^{(j)}(t,n)\mid g_{1},e_{2},n+1\rangle
+X_{4}^{(j)}(t,n)\mid g_{1},g_{2,}n+2\rangle \right], \quad
j=1,2,3 \label{new3}
\end{array}
\end{eqnarray}
\end{widetext}
where
\begin{widetext}
\begin{eqnarray}
\begin{array}{lr}
 X_{1}^{(1)}(t,n)=\mu^-_n\frac{\cos(t\Omega^+_n)}{2\Lambda_n}
+\mu^+_n\frac{\cos(t\Omega^-_n)}{2\Lambda_n},
\\
X_{2}^{(1)}(t,n)=\frac{-i\lambda_2\sqrt{n+1}}{2\Lambda_n}\{
[4\lambda_1^2(n+2)
 -\mu^-_n]\frac{\sin(t\Omega^+_n)}{\Omega^+_n}
-[4\lambda_1^2(n+2)
 -\mu^+_n]\frac{\sin(t\Omega^-_n)}{\Omega^-_n}\},
\\
X_{3}^{(1)}(t,n)=\frac{-i\lambda_1\sqrt{n+1}}{2\Lambda_n}\{
[4\lambda_2^2(n+2)
 -\mu^-_n]\frac{\sin(t\Omega^+_n)}{\Omega^+_n}
-[4\lambda_2^2(n+2)
 -\mu^+_n]\frac{\sin(t\Omega^-_n)}{\Omega^-_n}\},
\\
X_{4}^{(1)}(t,n)=\frac{2\lambda_1\lambda_2\sqrt{(n+2)(n+1)}}{\Lambda_n}[\cos(t\Omega^+_n)-
\cos(t\Omega^-_n)] ,\\
 X_{1}^{(2)}(t,n)=\frac{2\lambda_1\lambda_2\sqrt{n(n-1)}}{\Lambda_{n-2}}\left[
 \cos(t\Omega^+_{n-2})-\cos(t\Omega^-_{n-2})\right],
\\
X_{2}^{(2)}(t,n)=\frac{-i\lambda_1\sqrt{n}}{2\Lambda_{n-2}}\left\{
[4\lambda_2^2(n-1)
 +\mu^+_{n-2}]\frac{\sin(t\Omega^+_{n-2})}{\Omega^+_{n-2}}
-[4\lambda_2^2(n-1)
 +\mu^-_{n-2}]\frac{\sin(t\Omega^-_{n-2})}{\Omega^-_{n-2}}\right\},
\\
X_{3}^{(2)}(t,n)=\frac{-i\lambda_2\sqrt{n}}{2\Lambda_{n-2}}\left\{
[4\lambda_1^2(n-1)
 +\mu^+_{n-2}]\frac{\sin(t\Omega^+_{n-2})}{\Omega^+_{n-2}}
-[4\lambda_1^2(n-1)
 +\mu^-_{n-2}]\frac{\sin(t\Omega^-_{n-2})}{\Omega^-_{n-2}}\right\},
\\
X_{4}^{(2)}(t,n)=\frac{\mu^+_{n-2}}{2\Lambda_{n-2}}\cos(t\Omega^+_{n-2})-
\frac{\mu^-_{n-2}}{2\Lambda_{n-2}}\cos(t\Omega^-_{n-2}),\\
 X_{1}^{(3)}(t,n)=\frac{i\lambda_2\sqrt{n}}{2\Lambda_{n-1}}\left\{[\mu^-_{n-1}
 -4\lambda_1^2(n+1)]
 \frac{\sin(t\Omega^+_{n-1})}{\Omega^+_{n-1}}
 +[4\lambda_1^2(n+1)-\mu^+_{n-1}
]
 \frac{\sin(t\Omega^-_{n-1})}{\Omega^-_{n-1}}\right\},
 \\
X_{2}^{(3)}(t,n)=\frac{1}{2\Lambda_{n-1}}\left\{
[\lambda_1^2-\lambda_2^2+\Lambda_{n-1}]\cos(t\Omega^+_{n-1}) +
[\lambda_2^2-\lambda_1^2+\Lambda_{n-1}]\cos(t\Omega^-_{n-1})\right\},
\\
X_{3}^{(3)}(t,n)=\frac{\lambda_1\lambda_2(2n+1)}{\Lambda_{n-1}}[\cos(t\Omega^+_{n-1})
- \cos(t\Omega^-_{n-1})],
\\
 X_{4}^{(3)}(t,n)=\frac{i\lambda_1\sqrt{n+1}}{2\Lambda_{n-1}}\left\{[\mu^-_{n-1}+4\lambda_2^2n]
 \frac{\sin(t\Omega^-_{n-1})}{\Omega^-_{n-1}}-[\mu^+_{n-1}+4\lambda_2^2n]
 \frac{\sin(t\Omega^+_{n-1})}{\Omega^+_{n-1}}\right\}
  \label{new4}
\end{array}
\end{eqnarray}
\end{widetext}
and
\begin{widetext}
\begin{eqnarray}\label{insert1}
\begin{array}{lr}
\Lambda_n=\sqrt{(\lambda_1^2+\lambda_2^2)^2(2n+3)^2-4(\lambda_1^2-\lambda_2^2)^2(n+2)(n+1)},
\quad \mu^{\pm}_n=\lambda_1^2+\lambda_2^2\pm \Lambda_{n},
\\
\\
\Omega^\pm_n=\frac{1}{\sqrt{2}}\sqrt{(\lambda_1^2+\lambda_2^2)(2n+3)\pm
\Lambda_n}.
\end{array}
\end{eqnarray}
\end{widetext}

%%%%%%%%%%%%%%%%%%%%%%%%%%%%%%%%%%%%%%%%%%%%%%%%%%%%%%%%%%%%%%%%%%%
%%%%%%%%%%%%%%%%%%%%%%%%%%%%%%%%%%%%%%%%%%%%%%%%%%%%%%%%%%%%%%%%%%%%%
\begin{figure}[h]%
  \centering
  \subfigure[]{\includegraphics[width=8cm]{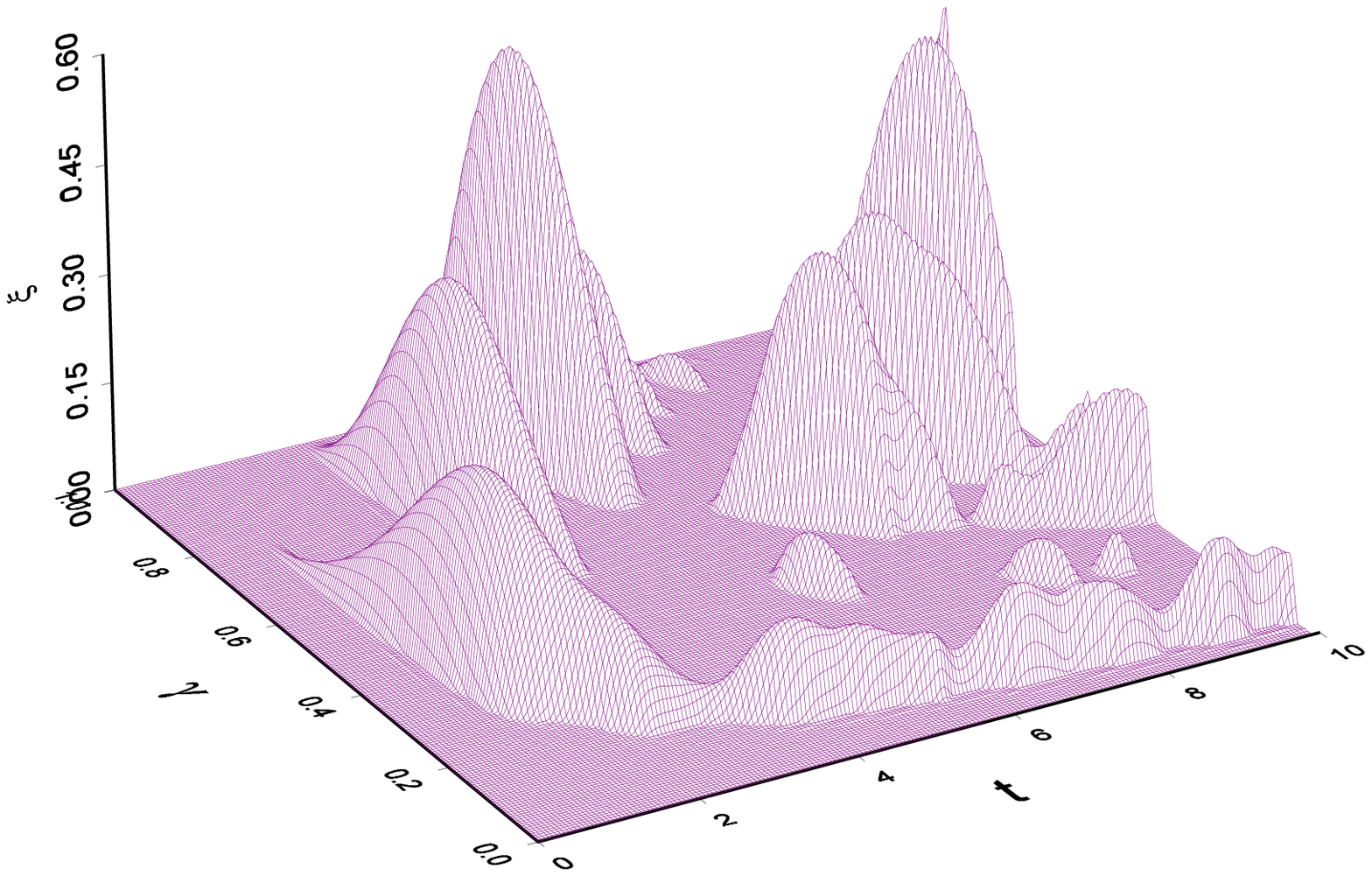}}
 \subfigure[]{\includegraphics[width=8cm]{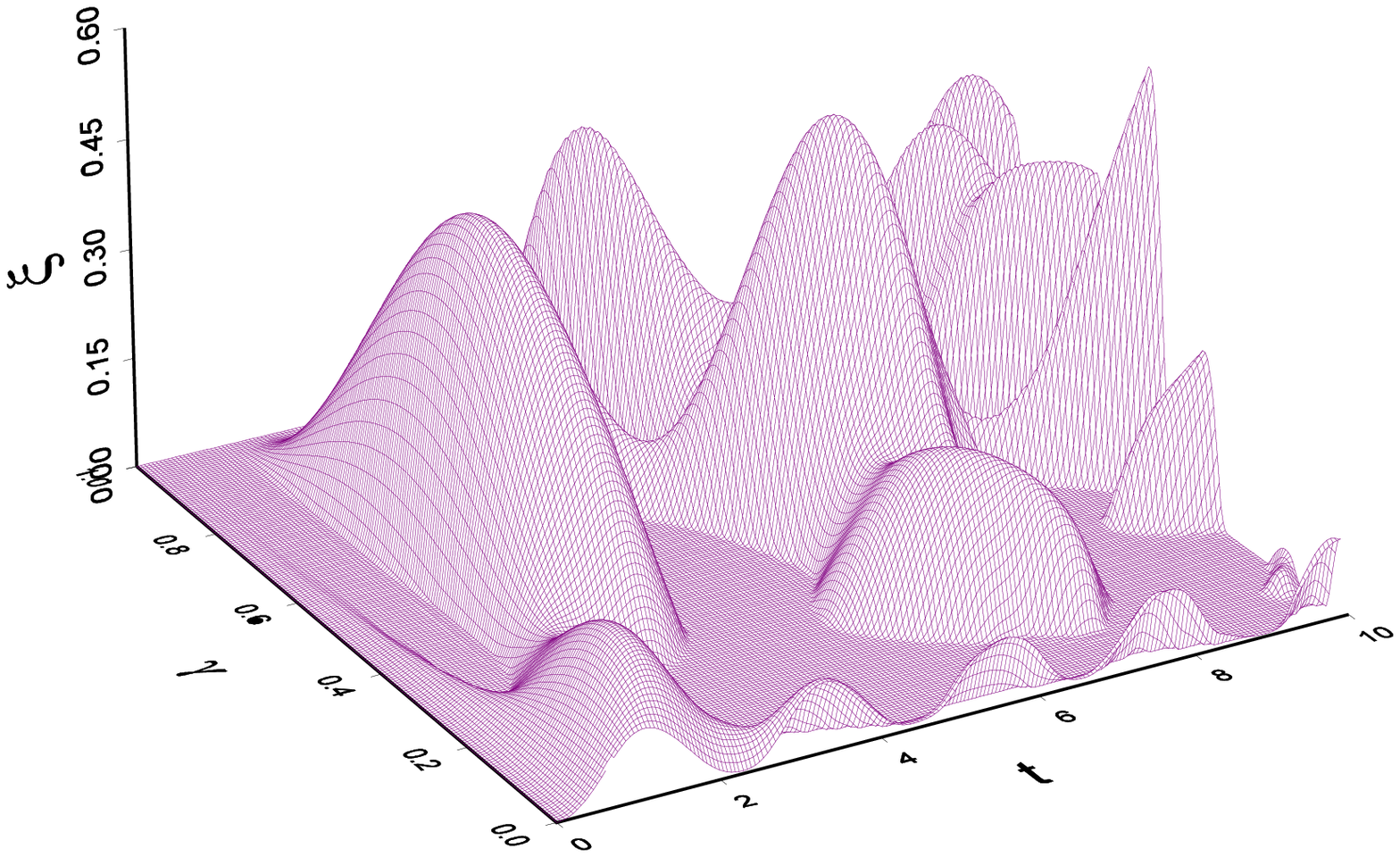}}
 % \subfigure[]{\includegraphics[width=8cm]{figc.eps}}
  %\subfigure[]{\includegraphics[width=8cm]{figd.eps}}
   \caption{
Atom-atom entanglement induced by the interaction with a thermal
field having $\bar{n}=1$ when the atoms are initially prepared in
$|e_1,e_2\rangle$ (a) and $|e_1,g_2\rangle$ (b).  }
  \label{fig1}
\end{figure}
%%%%%%%%%%%%%%%%%%%%%%%%%%%%%%%%%%%%%%%%%%%%%%%%%%%%%%%%%%%%%%%%%%%%%%%%%%
%%%%%%%%%%%%%%%%%%%%%%%%%%%%%%%%%%%%%%%%%%%%%%%%%%%%%%%%%%%%%%%%%%%%%%%%%

The dynamical density  matrix of the whole system can be obtained
by   substituting (\ref{new3})--(\ref{insert1})  into (\ref{1.7})
and then carrying out the integration with respect to $\phi$. This
treatment provides the  exact solution of the problem. Moreover,
it is easier and better than solving the Liouville equation
numerically \cite{lee1}. To evaluate the density matrix of the two
qubits, i.e. $\hat{\rho}_a(t)$, one has to trace over the field
variables the  density matrix of the system as:
\begin{equation}\label{new5}
    \hat{\rho}_a(t)=B_{e_1e_2}|e_1,e_2\rangle \langle e_1,e_2|+B_{g_1g_2}|g_1,g_2\rangle \langle
    g_1,g_2|+(B_{e_1g_2}|e_1,g_2\rangle+B_{g_1e_2}|g_1,e_2\rangle)(\langle e_1,g_2|B^*_{e_1g_2}+\langle
    g_1,e_2|B^*_{g_1e_2}),
\end{equation}
where
\begin{widetext}
\begin{eqnarray}\label{insert2}
\begin{array}{lr}
B_{e_1e_2}=\sum\limits_{n=0}^{\infty }P(n)\left[
|X_{1}^{(1)}(t,n)|^2\sin^2\theta\cos^2\vartheta+|X_{1}^{(2)}(t,n)|^2\sin^2\theta\sin^2\vartheta+
|X_{1}^{(3)}(t,n)|^2\cos^2\theta\right],
\\
B_{g_1g_2}=\sum\limits_{n=0}^{\infty }P(n)\left[
|X_{4}^{(1)}(t,n)|^2\sin^2\theta\cos^2\vartheta+|X_{4}^{(2)}(t,n)|^2\sin^2\theta\sin^2\vartheta+
|X_{4}^{(3)}(t,n)|^2\cos^2\theta\right],
\\
B_{e_1g_2}=\sum\limits_{n=0}^{\infty }P(n)\left[
X_{2}^{(1)}(t,n)X_{3}^{(1)*}(t,n)\sin^2\theta\cos^2\vartheta\right.\\
+\left.
X_{2}^{(2)}(t,n)X_{3}^{(2)*}(t,n)\sin^2\theta\sin^2\vartheta+
X_{2}^{(3)}(t,n)X_{3}^{(3)*}(t,n)\cos^2\theta\right],\quad
B_{g_1e_2}=B_{e_1g_2}^*.
\end{array}
\end{eqnarray}
\end{widetext}
For the best of our knowledge this is the first time the explicit
form of the density matrix of this system to be presented. This
reflects the power of the approach. The density matrix
(\ref{new5}) tends to that of the symmetric case \cite{lee2},
which was obtained by the Kraus representation, by simply setting
$\lambda_1=\lambda_2=\lambda$.

To verify the validity of the approach we comment on the
entanglement of the two qubits controlled by (\ref{new5}).

In this regard, we  use the negative values of the partial
transposition \cite{lee}, which is frequently used for such
systems \cite{amesen,lee2}, and is defined as:
\begin{equation}\label{masu}
    \xi=-2\sum_j \gamma_j^-,
\end{equation}
where $\gamma_j^-$ are the negative eigenvalues of the partial
transposition of $\hat{\rho}_a(t)$. The entanglement measure
ranges between  $\xi=0$ for    separable qubits  and $\xi=1$ for
maximally entangled qubits.
 For the density matrix (\ref{new5}) there is only
one eigenvalue of those of the partial transposition matrix, which
can approach negative values, having the form:
\begin{equation}\label{negativ}
2
\gamma_j^-=B_{e_1e_2}+B_{g_1g_2}-\sqrt{(B_{e_1e_2}-B_{g_1g_2})^2+4|B_{g_1e_2}|^2}.
\end{equation}
By means of (\ref{new4})--(\ref{negativ})
 we have obtained the results of
 \cite{lee2} for the same values of the interaction
parameters.  Now it is therefore reasonable to make a comparison
with the results of the general case, which has been numerically
presented
 in \cite{lee1}. In that article the analysis has been
 confined  to such forms of
 the coupling constants  $\lambda_1=1+\gamma$ and
$\lambda_2=1-\gamma$, where $\gamma$ (with $0\leq \gamma\leq 1$)
is the relative difference between the two atomic couplings. This
means that the strength of the interaction of  one of the
bipartites (atom-field) is increasing, while  the other is
 decreasing simultaneously. As an example, in Figs. 1 we plot the quantity
$\xi$ for the same values of the interaction parameters as those
of Figs. 2 in \cite{lee1}, which were given for  the concurrence.
It is obvious that our figures and those in \cite{lee1} are
identical even though they represent different measures. This
clearly confirm the validity of the approach. Obviously, the
analytical treatments are in general better than the numerical
ones as they can provide us
 some analytical facts about the system.
For instance, the condition of involving the expression
(\ref{negativ})  negative values  is:
\begin{equation}\label{final11}
\Upsilon=B_{e_1e_2}B_{g_1g_2}-|B_{g_1e_2}|^2<0.
\end{equation}
From (\ref{new4})--(\ref{insert2}), when $n=0$ and the two atoms
are in $|e_1,g_2\rangle$, the inequality (\ref{final11}) can be
simplified as:
\begin{equation}\label{final1}
\Upsilon=-\frac{\lambda_1\lambda_2}{(\lambda_1^2+\lambda_2^2)^2}
[\cos(t\Omega^+_{-1})-1][\lambda_1^2\cos(t\Omega^+_{-1})+\lambda_2^2]<0.
\end{equation}
It is evident that the expression (\ref{final1}) gives
$\Upsilon=\sin^2(t\Omega^+_{-1})/4$ for the symmetric case. This
means that the symmetric case (with $\bar{n}=0$) cannot generate
any entanglement between the two atoms, which are initially in
$|e_1,g_2\rangle$. Nevertheless, the asymmetric case can generate
entanglement
 for certain values of the coupling constants, in particular,
for those satisfying  the inequality
$\cos(t\Omega^+_{-1})<-\lambda_2^2/\lambda_1^2$. Furthermore, for
 the atoms, which  are initially in $|g_1,g_2\rangle$ and $n=0$,
one can easily prove that $X_j^{(3)}(t,0)=0$. Thus, the condition
(\ref{final11}) is not fulfilled, i.e. entanglement cannot be
established in this case.

It is worth  prompting   that we have aimed by the above
discussion to justify the validity of the approach not to repeat
the study of the entanglement  of the two qubits. So that we have
selected few cases for the sake of comparison only. Nevertheless,
from the treatment we have performed, which was not presented here
for the sake of brevity, and the other treatments given in
\cite{amesen,lee1,lee2} we can conclude that a highly chaotic
state in an infinite-dimensional Hilbert space can entangle two
qubits depending on the type of their interaction with the field
as well as the initial conditions of the system. The amounts of
entanglement generated by the asymmetric case are much greater
than those generated by the symmetric one.

In conclusion, we have developed, for the first time, a simple
analytical  approach for solving  unitary system when the initial
field, as one of its components, is in the mixed state, e.g.
thermal light. The approach is applicable  only when the
 Schr\"{o}dinger's equation of the system is solvable for
   any initial arbitrary  state. We have verified  the validity of the approach
      for some selective cases for the entanglement of two
      qubits. The approach enables us to check  the results, which
      have been numerically
obtained earlier \cite{lee1}.
   As a final note, we believe that
our approach  represents a powerful tool to solve
 such type of  problems and
may impact other applications in the quantum theory.

%%%%%%%%%%%%%%%%%%%%%%%%%%%%%%%%%%%%%%%%%%%%%%%%%%%%%%%%%%%%
\section*{ Acknowledgment}
%%%%%%%%%%%%%%%%%%%%%%%%%%%%%%%%%%%%%%%%%%%%%%%%%%%%%%%%%%%%%%

 The author would like to thank Professors Z. Ficek and M. S. Abdalla
 for the critical reading of the manuscript.

\end{document}